\documentclass[runningheads]{llncs}

\usepackage{xcolor}
\usepackage{xspace}
\usepackage{amsfonts}
\usepackage{tikz}
\usepackage{pgfplots}
\usepackage{listings,chngcntr}
\usepackage{placeins}
\usepackage{hyperref}
\usepackage{multirow}
\usepackage{algorithm}
\usepackage{algorithmic}
\usepackage{graphicx}

\usepackage[firstpage]{draftwatermark}
\usepackage[caption=false]{subfig}
\usepackage[english]{babel}
\usepackage[utf8]{inputenc}

\pgfplotsset{compat=1.6}
\usetikzlibrary{patterns,positioning,shapes.geometric}

\hypersetup{colorlinks=false}
\newcommand{\algref}[1]{Alg.~\ref{Alg:#1}}
\newcommand{\coderef}[1]{Listing~\ref{Cd:#1}}
\newcommand{\figref}[1]{Fig.~\ref{Fi:#1}}
\newcommand{\sectref}[1]{Section~\ref{Se:#1}}
\newcommand{\tableref}[1]{Tab.~\ref{Ta:#1}}
\newcommand{\algline}[1]{(Line~\ref{algte:#1})}
\newcommand{\alglines}[2]{(Line~\ref{algte:#1}--\ref{algte:#2})}
\newcommand{\codeline}[1]{Line~\ref{Line:#1}}

\newcommand{\threelines}[3]{Lines~\ref{Line:#1}, ~\ref{Line:#2}, and ~\ref{Line:#3}}

\definecolor{gray(x11gray)}{rgb}{0.6, 0.6, 0.6}
\newcommand{\highlightcode}[1]{\makebox[0pt][l]{\color{lightgray}\rule[-3pt]{#1\linewidth}{10pt}}}
\newcommand{\gridcode}[1]{\makebox[0pt][l]{\color{gray(x11gray)}\rule[-3pt]{#1\linewidth}{10pt}}}

\newcommand{\True}{\ensuremath{\mathit{true}}\xspace}
\newcommand{\False}{\ensuremath{\mathit{false}}\xspace}

\newcommand{\inangleb}[1]{ \ensuremath{\langle #1 \rangle}}

\newcommand{\toolname}{GPURepair\xspace}
\newcommand{\verifiername}{GPUVerify\xspace}
\newcommand{\autosyncname}{AutoSync\xspace}

\newcommand{\tool}{\textsf{GPURepair}\xspace}
\newcommand{\verifier}{\textsf{GPUVerify}\xspace}
\newcommand{\autosync}{\textsf{AutoSync}\xspace}
\newcommand{\zth}{\textsf{Z3}\xspace}
\newcommand{\boogie}{\textsf{Boogie}\xspace}

\newcommand{\app}[1]{\textsf{#1}\xspace}

\newcommand{\emp}[1]{\textsf{#1}}
\newcommand{\TT}[1]{{\tt #1}}
\newcommand{\EM}[1]{{\em #1}}
\newcommand{\MB}[1]{\mathbf{#1}}

\newcommand{\sjorcidlogo}{\href{https://orcid.org/0000-0001-8070-1525}{\protect\includegraphics[scale=0.4]{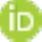}}}
\newcommand{\gmorcidlogo}{\href{https://orcid.org/0000-0002-3632-7801}{\protect\includegraphics[scale=0.4]{images/orcid.pdf}}}

\begin{document}
\title{\toolname: Automated Repair of GPU Kernels}

\author{Saurabh Joshi \textsuperscript{\sjorcidlogo} \and
Gautam Muduganti \textsuperscript{\gmorcidlogo}
\thanks{The author names are in alphabetical order.}}

\authorrunning{Joshi and Muduganti}

\institute{Indian Institute of Technology Hyderabad, India
\email{\{sbjoshi,cs17resch01003\}@iith.ac.in}}

\SetWatermarkText{\raisebox{11.4cm}{
	\includegraphics[scale=0.9]{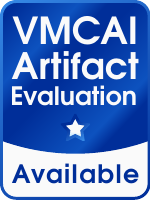} \hspace{0.75cm} \includegraphics[scale=0.9]{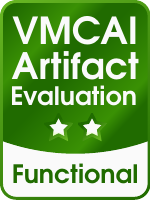} \hspace{0.75cm} \includegraphics[scale=0.9]{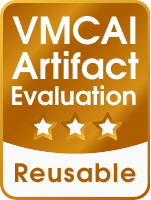}
}}
\SetWatermarkAngle{0}

\maketitle
\vspace{1.35cm}

\begin{abstract}
This paper presents a tool for repairing errors in GPU kernels written in CUDA or OpenCL due to data races and barrier divergence. Our novel extension to prior work can also remove barriers that are deemed unnecessary for correctness. We implement these ideas in our tool called \tool, which uses \verifier as the verification oracle for GPU kernels. We also extend \verifier to support CUDA Cooperative Groups, allowing \tool to perform inter-block synchronization for CUDA kernels. To the best of our knowledge, \tool is the only tool that can propose a fix for intra-block data races and barrier divergence errors for both CUDA and OpenCL kernels and the only tool that fixes inter-block data races for CUDA kernels. We perform extensive experiments on about 750 kernels and provide a comparison with prior work. We demonstrate the superiority of \tool through its capability to fix more kernels and its unique ability to remove redundant barriers and handle inter-block data races.
\end{abstract}

\keywords{GPU, Verification, Automated Repair, CUDA, OpenCL}

\section{Introduction}
\label{Se:introduction}
The part of the program that runs on the GPU (Graphics Processing Unit) is referred to as a \EM{kernel}. Given that multiple cores of the GPU may execute the kernel in parallel, data races and barrier divergence are frequently the cause of several errors that occur in practice. Identifying and repairing these errors early in the development cycle can have a tremendous positive financial impact \cite{boehm1988understanding}.

In CUDA, a grid consists of blocks, and a block consists of threads. Consider the CUDA kernel in \coderef{race} without the highlighted line. There is a data race on accesses of the shared array \texttt{A}. The race can be avoided by introducing a barrier (\TT{\_\_syncthreads()}) in the kernel at line \ref{codeline:race:barrier} in \coderef{race}. This block-level barrier enforces that all threads in a block reach it before any of them can proceed further. A grid-level barrier behaves similarly for the entire grid.

In \coderef{divergence}, only the threads with an even thread id will reach the barrier. As the threads within a block execute in a \EM{lock-step} manner, this will result in a deadlock as threads with odd ids will never be able to reach the barrier at \codeline{di:ba}. This problem is known as \EM{barrier divergence}.

\noindent\begin{tabular}{ll}
	\begin{minipage}{0.45\textwidth}
\begin{lstlisting}[caption=Kernel with Data Race, label={Cd:race}, captionpos=b, xleftmargin=2em, escapechar=\^]
__global__ void race (int* A) {
  int temp = A[threadIdx.x+1];
  ^\highlightcode{0.5}^__syncthreads(); ^\label{codeline:race:barrier}^
  A[threadIdx.x] = temp;
}
\end{lstlisting}

	\end{minipage} &
	\begin{minipage}{0.55\textwidth}
\begin{lstlisting}[caption=Kernel with Barrier Divergence, label={Cd:divergence}, captionpos=b, xleftmargin=3em, escapechar=\^]
__global__ void race (int* A) {
  if (threadIdx.x % 2 == 0) {
    int temp = A[threadIdx.x+1];
    __syncthreads(); ^\label{Line:di:ba}^
    A[threadIdx.x] = temp;
  }
}
\end{lstlisting}

	\end{minipage}
\end{tabular}

This tool paper makes the following contributions:

\begin{itemize}
\item We extend the underlying technique behind \autosync \cite{anand2018automatic} to provide barrier placements that avoid barrier divergence in addition to data races. Our novel extension may also suggest removing barriers inserted by the programmer if deemed unnecessary, which might help enhance the performance of the input GPU kernel.
\item We implement our technique in our tool \tool, which is built on top of the \verifier \cite{betts2012gpuverify} framework and uses \verifier as an oracle. To the best of our knowledge, ours is the only technique and tool that can propose a fix for both CUDA and OpenCL GPU kernels. Another unique feature of \tool is its ability to fix kernels that have inter-block data races.
\item \app{Bugle} is the component of the \verifier toolchain that translates LLVM bitcode to \boogie. We have enhanced it with the ability to translate barrier synchronization statements from the CUDA Cooperative Groups API to \boogie. We have also extended \verifier with the semantics to support grid-level barriers. Using these enhancements, \tool proposes fixes for inter-block data races.
\item We perform an extensive experimental evaluation on $748$ GPU kernels written in CUDA and OpenCL. We compare \tool against \autosync, which is the only other tool known that attempts to repair CUDA kernels containing data races.
\end{itemize}

\section{\tool}
\label{Se:gpurepair}

\subsection{\toolname architecture and workflow}
\label{Se:workflow}
\begin{figure}[t]
\centering
\includegraphics[scale=0.25]{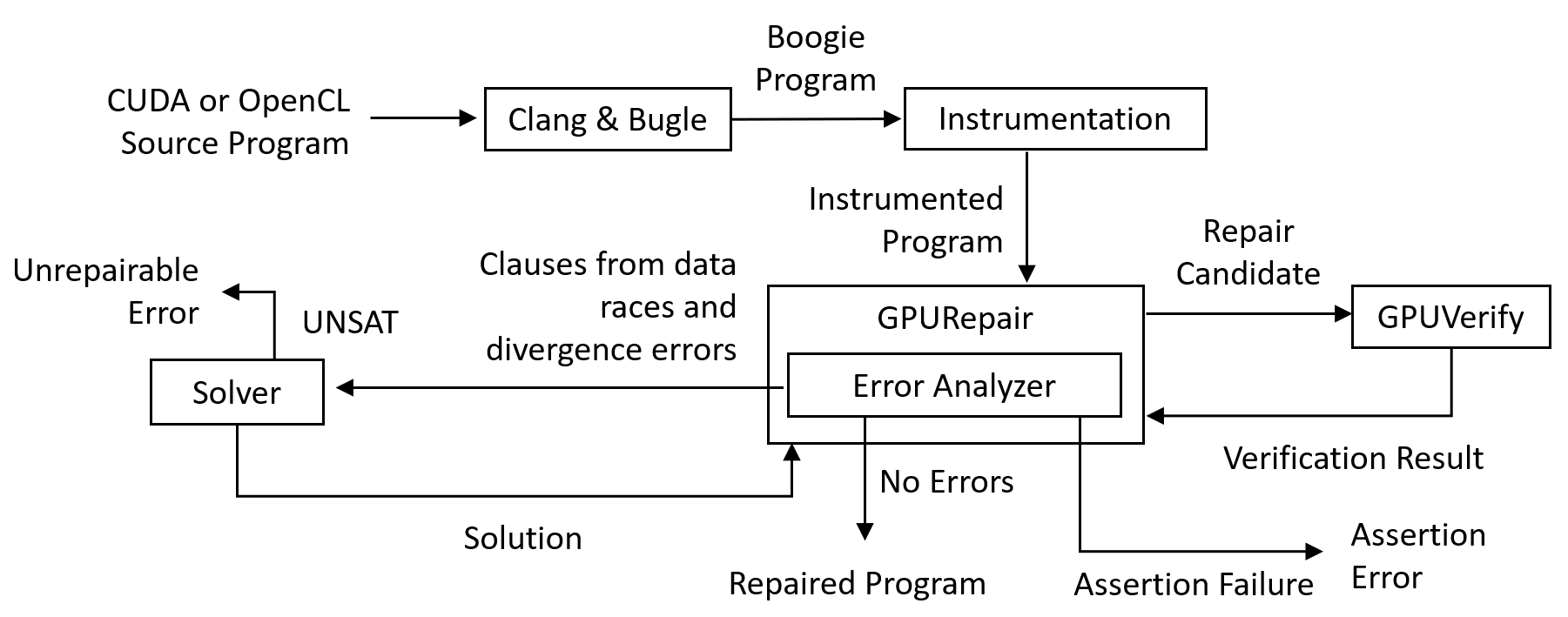}
\caption{\toolname Workflow}
\label{Fi:repairer_workflow}
\end{figure}

The implementation of \tool builds on top of \verifier, as depicted in \figref{repairer_workflow}. In addition to the instrumentation done by \verifier to enable verification of GPU kernels, \tool adds the instrumentation necessary to impose constraints on the program behavior. On each iteration, \tool calls \verifier with a proposed solution to check if the program is repaired. If the program is not repaired, it calls the \EM{Solver} with the constraints generated from the errors seen so far to obtain a candidate solution suggesting which barriers need to be enabled/disabled in the program. If the program can be repaired, \tool generates the \boogie representation of the fixed program and a summary file. The summary file contains the changes that have to be made to fix the program along with the source location details of the original CUDA/OpenCL input kernel. The technique behind \tool can, in principle, use any verifier for GPU programs as an oracle.

\subsection{Instrumentation}
\label{Se:instrumentation}

\verifier uses a pair of distinct non-deterministically chosen threads for analysis instead of modeling all the threads in the kernel. This two-thread abstraction is used to prove that a kernel is race-free and divergence-free. Details of this abstraction are available in \cite{betts2012gpuverify} and beyond the scope of this paper. \verifier models barriers by resetting read/write flags of shared arrays for the two threads used by the two-thread abstraction if they belong to the same block. We extend this to support grid-level barriers for the repair of inter-block data races. A grid-level barrier is modeled by performing a reset even when the two threads do not belong to the same block.

Since \tool attempts to fix errors caused only due to data races or barrier divergence, it proposes a solution that only involves removing existing barriers or adding new ones. Therefore, the instrumentation stage of \tool introduces barriers guarded with Boolean variables, referred to as \EM{barrier variables}. The value of a barrier variable acts as a switch to enable or disable the barrier. A barrier guarded by a barrier variable is referred to as an \EM{instrumented barrier}. Consider the kernel in \coderef{kernel_example} without the highlighted lines. This kernel has a data race. The instrumentation process adds an instrumented barrier before a shared variable is either read or written. Function calls involving a shared variable are also taken into consideration, and an instrumented barrier is added before the invocation. Pre-existing barriers in the programs are also guarded with barrier variables to explore if they can be removed without introducing data races.

In addition, the control flow graph (CFG) is analyzed for branch statements and loops to handle scenarios where barriers, if inserted right before the read/write to a shared variable, may introduce barrier divergence. For example, in \coderef{kernel_example}, the instrumentation process mentioned above would insert the instrumented barriers right before \codeline{ke:read} and \codeline{ke:write}. However, the solution to this program would be a barrier before the \TT{if} block at \codeline{ke:joinpoint}. The instrumentation process takes these scenarios under consideration by inserting instrumented barriers at the scope boundaries such as entry points of branch statements, loop-heads, and function calls. After instrumentation, the highlighted lines at \threelines{ke:ba}{ke:bb}{ke:bc} in \coderef{kernel_example} are added. For CUDA kernels, if instrumentation of grid-level barriers is enabled, the highlighted lines at \threelines{ke:bd}{ke:be}{ke:bf} are also added.

\vspace{-0.5cm}
\noindent\begin{tabular}{ll}
	\begin{minipage}{0.5\textwidth}
\begin{algorithm}[H]
\caption{The Repair Algorithm}
\label{Alg:gpurepair}
\begin{scriptsize}
\begin{algorithmic}[1]
\STATE \emp{Input:} Instrumented Program $P$
\STATE \emp{Output:} Repaired Program $P_\varphi$

\STATE $\varphi := true$ \label{algte:constraint}

\LOOP \label{algte:startloop}
\STATE $\inangleb{res,sol} := Solve(\varphi)$ \label{algte:solve}
\IF {$res = UNSAT$} \label{algte:unsat}
\PRINT \emp{Error:} Program cannot be repaired \label{algte:unsaterror}
\RETURN \emp{errorcode}
\ENDIF
\STATE $\inangleb{result, \pi} := Verify(P_{sol})$ \label{algte:verify}

\IF {$result = SAFE$} \label{algte:repaired}
\STATE \emp{break}
\ENDIF

\IF {$result \neq RACE\ \&\&\ result \neq DIVERGENCE$} \label{algte:asserterror}
\PRINT \emp{Error:} Program cannot be repaired
\RETURN \emp{errorcode}
\ENDIF

\STATE $c := GenerateClause(\pi)$ \label{algte:generateclause}
\STATE $\varphi := \varphi \cup \{c\}$ \label{algte:addclause}
\ENDLOOP \label{algte:endloop}

\RETURN $P_{sol}$ \label{algte:returnsol}
\end{algorithmic}
\end{scriptsize}

\end{algorithm}

	\end{minipage}
	\begin{minipage}{0.5\textwidth}
\vspace{1cm}
\begin{lstlisting}[caption=Example CUDA Kernel, label={Cd:kernel_example}, captionpos=b, xleftmargin=3em, escapechar=\^]
bool b1, b2, b3, b4, b5, b6;
__device__ void write(int* A, int idx) {
  A[idx] = 50; ^\label{Line:ke:inside}^
}

__global__ void race(int* A) {
  auto g = this_grid();

  ^\highlightcode{0.83}^if(b1) { __syncthreads(); } ^\label{Line:ke:ba}^
  ^\gridcode{0.83}^if(b4) { g.sync(); } ^\label{Line:ke:bd}^
  int temp = A[threadIdx.x+1]; ^\label{Line:ke:read}^
  
  ^\highlightcode{0.83}^if(b2) { __syncthreads(); } ^\label{Line:ke:bb}^
  ^\gridcode{0.83}^if(b5) { g.sync(); } ^\label{Line:ke:be}^
  if (temp < 50) { ^\label{Line:ke:joinpoint}^
    ^\highlightcode{0.83}^if(b3) { __syncthreads(); } ^\label{Line:ke:bc}^
    ^\gridcode{0.83}^if(b6) { g.sync(); } ^\label{Line:ke:bf}^
    write(A, threadIdx.x); ^\label{Line:ke:write}^
  }
}
\end{lstlisting}

	\end{minipage}
\end{tabular}
\vspace{0.2cm}

In this example, variables \texttt{b1,\ldots,b6} are initially unconstrained. Their values are constrained by \tool iteratively to avoid data races or barrier divergence during the repair process. The repair process also assigns weights to these barrier variables such that introducing barriers at the block-level is preferred over the grid-level. This is done because grid-level barriers have a higher performance penalty \cite{zhang2020study}. For the same reason, barriers nested within loops are less preferred. Although the examples in this section are in CUDA, it should be noted that the actual working of this stage happens on the \boogie program generated by \app{Bugle} to make \tool agnostic to the front-end language (i.e., CUDA or OpenCL).

\subsection{Preliminaries}
\label{Se:preliminaries}
Let $P$ be the given input GPU kernel after instrumentation, as described in \sectref{instrumentation}. Let $\{ b_1,\dots,b_m \}$ be the barrier variables introduced as a part of the instrumentation process. Given a formula $\varphi$ over $b_i$'s, let $P_{\varphi}$ denote the instrumented kernel $P$ with values of $b_i$'s constrained to obey $\varphi$.

A clause $c$ is called a \EM{positive} (resp.\EM{ negative}) \EM{monotone clause} if it has literals with only positive (resp. negative) polarity (e.g., $b_1 \vee b_5 \vee b_{11}$). From now on, we may also denote a clause as a set of literals with disjunctions among the set elements being implicit. A formula or a constraint $\varphi$ is a set of clauses with conjunction being implicit among the set elements. Note that a formula $\varphi$ consisting of only positive monotone clauses or only negative monotone clauses is always satisfiable. Let $\varphi^+$ (resp. $\varphi^-$) denote the set of positive (resp. negative) monotone clauses belonging to $\varphi$.

Let $C$ be a set consisting of non-empty sets $S_1, \dots, S_n$. The set $\mathcal{H}$ is called a \emph{hitting-set} (HS) of $C$ if:

\[ \forall_{S_i \in C} \mathcal{H} \cap S_i \neq \emptyset \]

$\mathcal{H}$ is called a \emph{minimal-hitting-set} ($mhs$) if any proper subset of $\mathcal{H}$ is not a hitting-set. $\mathcal{H}$ is called a \EM{Minimum-Hitting-Set} ($MHS$) of $C$ if no smaller hitting set exists for $C$. Note that a collection $C$ may have multiple $mhs$ and multiple $MHS$.

Since we also consider a formula $\varphi$ as a set of clauses, we shall abuse the notation and use $mhs(\varphi^+)$ to denote the set of literals that constitutes the minimal-hitting-set of $\varphi^+$.

Maximum satisfiability ($MaxSAT$) is an optimization version of the $SAT$ problem where the goal is to find an assignment that maximizes the number of clauses that are satisfied in a formula. In partial $MaxSAT$ (PMS), given a set of hard clauses ($\varphi_h$) and a set of soft clauses ($\varphi_s$), the goal is to find an assignment that satisfies all the clauses of $\varphi_h$ while maximizing the number of soft clauses being satisfied. The weighted partial $MaxSAT$ (WPMS) problem asks to satisfy all the hard clauses while maximizing the sum of the weights of the satisfied soft clauses. In WPMS, positive weights are associated with each soft clause.

\subsection{The Repair Algorithm}
\label{Se:repair_algorithm}

\algref{gpurepair} depicts the repair technique behind \tool at a high level. It is very similar to the algorithm presented in \autosync \cite{anand2018automatic}.

Initially, all the barrier variables are unconstrained \algline{constraint}, giving the verifier the freedom to set them to any value that leads to an error. Then, \algref{gpurepair} iteratively calls the verifier \alglines{startloop}{endloop} until it either finds a solution or determines that it cannot repair the program. A call to the verifier \algline{verify} returns an error trace $\pi$ along with the type of the error being captured in $result$. If the verifier could not find an error \algline{repaired} with the proposed solution $sol$, then the algorithm exits the loop, and the instrumented \boogie program constrained with $sol$ is returned \algline{returnsol}. If the verifier returned with an error that is neither a data race nor a barrier divergence \algline{asserterror}, then \algref{gpurepair} terminates with an error stating that it cannot repair the program. Here, we are operating under the assumption that inserting an extraneous barrier may only introduce a barrier divergence error, and removing a programmer-inserted barrier may only cause a data race error.

If the verifier returns with a data race error, then the error trace $\pi$ would tell us which set of barriers were disabled (i.e., the corresponding barrier variables were set to \False by the verifier). Let the set of barrier variables that were set to \False in $\pi$ be $b_{i_1},\dots,b_{i_d}$. To avoid the same error trace $\pi$ we need to add a constraint represented as a clause $c$, $(b_{i_1}\vee,\dots,\vee b_{i_d})$, which is generated by $GenerateClause(\pi)$ \algline{generateclause}. Note that such a clause generated from a data race (respectively barrier divergence) error always has only positive (resp. negative) literals. This newly generated clause is added to the constraint $\varphi$ \algline{addclause}, which consists of one clause per error trace. We need to check if $\varphi$ is satisfiable \algline{solve}. If it is not satisfiable \algline{unsat}, it indicates that there is no assignment to barrier variables that avoids all previously seen traces. \algref{gpurepair} quits with an error \algline{unsaterror} in this case. If $\varphi$ is satisfiable, then the $Solve$ method proposes a solution $sol$ \algline{solve}, which is essentially an assignment to some of the barrier variables. We use two different ways to compute $sol$ from $\varphi$. The first method (the $MaxSAT$ strategy) uses a $MaxSAT$ solver to minimize the number of barrier variables being set to $\True$ in $sol$ at each iteration. This is done by solving a partial $MaxSAT$ problem with $\varphi$ as hard clauses and $\{\neg b_1,\dots,\neg b_m\}$ as soft clauses. The second method (the $mhs$ strategy) is to compute a minimal-hitting-set ($mhs$) over $\varphi^+$ using a polynomial-time greedy algorithm \cite{johnson1974approximation} at each iteration to attempt to minimize the number of $b_i$'s being set to $\True$. In this strategy, a single query to a $MaxSAT$ solver is needed to ensure that the number of $b_i$'s being set to $\True$ is the minimum.

A similar approach has been used previously in other works \cite{joshi2014automatically,abdulla2012tacas,joshi2015property}. It should be noted that the clauses generated in these works are all positive monotone clauses (clauses with only positive literals). In contrast, the clauses generated by \algref{gpurepair} could be a mix of positive monotone clauses and negative monotone clauses. Because of this added complication, the approach of using the $mhs$ is not complete. There could be a scenario where the $mhs$ could come up with a solution that causes the negative monotone clauses to be unsatisfiable. Consider an example with the following clauses: \{ ${b1 \vee b3}$, ${b1 \vee b4}$, ${b2 \vee b5}$, ${b2 \vee b6}$, ${\neg b1 \vee \neg b2}$ \}. The $mhs(\varphi^+)$ would give us $b1$ and $b2$, which would cause the clause ${\neg b1 \vee \neg b2}$ to be unsatisfiable. To overcome this, \tool falls back to the $MaxSAT$ solver whenever the $mhs$ approach results in unsatisfiability.

A barrier inside a loop can pose a heavier performance penalty as opposed to a barrier that is not nested inside a loop. Similarly, inter-block synchronization is more expensive compared to intra-block synchronization \cite{zhang2020study}. In principle, different barrier variables can be given different weights based on loop nesting or profiling information. Then, instead of minimizing the number of barriers, one may want to have barrier placements that minimize the sum of the weights of the enabled barriers. This can easily be achieved by posing this as a weighted $mhs$ or a weighted partial $MaxSAT$ (WPMS) problem. The weight of a barrier is computed as $((gw*gb)+lw^{ld})$ where $gw$ is the penalty for a grid-level barrier, $gb$ is $0$ for block-level barriers and $1$ for grid-level barriers, $lw$ is the penalty for a barrier that is inside a loop, and $ld$ is the loop-nesting depth of the barrier.

\section{Related Work}
\label{Se:relatedwork}
The verification of GPU programs has been an active area of research for quite a while. \verifier \cite{betts2012gpuverify,betts2015design} defines an operational semantics for GPU kernels called synchronous, delayed visibility (SDV) semantics, which mimics the execution of a kernel by multiple groups of threads and uses this to identify data races and barrier divergence. \app{ESBMC-GPU} \cite{monteiro2018esbmc} extends the ESBMC model checker by generating an abstract representation of the CUDA libraries and verifies kernels by checking specific desired properties. \app{VerCors} \cite{blom2014specification,amighi2015specification} builds on permission-based separation logic and extends it to verify GPU programs. \app{PUG} \cite{li2010scalable} is a symbolic verifier that uses SMT solvers to identify bugs in CUDA kernels. Contrary to most of the other verifiers which use static analysis for verification, \app{GKLEE} \cite{li2012gklee} uses concolic verification to identify concrete witnesses for reported bugs.

Automatic program repair is another active area of research that ties-in quite closely with our work. There have been several research efforts in the past for repairing sequential programs \cite{chandra2011angelic,griesmayer2006repair,jobstmann2005program,malik2011constraint} as well as concurrent programs \cite{deshmukh2010logical,jin2011automated,muzahid2010atomtracker,vcerny2011quantitative,vechev2010abstraction,joshi2015property,joshi2014automatically}. The work done in \cite{deshmukh2010logical,vechev2010abstraction,joshi2015property,joshi2014automatically} is very similar to the approach that we take in this paper, where the source code is instrumented, and the repair technique uses the error traces obtained from a verifier to fix the program.

To the best of our knowledge, apart from \tool, \autosync \cite{anand2018automatic} is the only tool that tries to repair a GPU program. \autosync takes a CUDA program \EM{without any barriers} and introduces barriers at appropriate locations to remove data race errors.

\subsection{Comparison with \autosyncname}
\label{Se:comparison}

\noindent\textbf{Repair on Source Code vs. Repair on Boogie Code:} \autosync and \tool use different types of inputs. \autosync uses a CUDA program as its input for the repair process, whereas \tool uses the \boogie program generated from \app{Bugle}. \tool is agnostic to the front-end language, allowing it to handle both CUDA and OpenCL. \autosync, on the other hand, directly takes and manipulates the source code, which makes it fragile to syntactic changes and limits its capabilities to CUDA only.

Consider the kernel in \coderef{kernel_example}. When function inlining is enabled, \verifier can accurately identify that \codeline{ke:inside} and \codeline{ke:read} cause a read-write race through the function call at \codeline{ke:write}. It reports the line information by specifying that the lines inside the functions cause the read-write race and also explicitly specifies the lines from where these functions are invoked. \autosync, however, does not process this information correctly and ends up with a code error. \tool takes these cases into account and has the ability to place the barrier precisely between the function invocations.

\noindent\textbf{Barrier Placement vs. Instrumentation:} \autosync uses \verifier to identify the lines of code that cause the data race and tries to insert a barrier between these lines of code. In contrast, \tool inserts instrumented barriers at various locations in the intermediate \boogie code based on the usage of global variables and uses the trace information provided in errors to enable/disable barriers. This instrumentation gives \tool the ability to remove programmer-inserted barriers that are deemed unnecessary as well as repair errors caused by barrier divergence or data races that require a grid-level barrier. This feature is exclusive to \tool.

Consider the statement \TT{A[idx] = A[idx + 1]}, that has a read-write race occurring in a single line of code. Because \autosync uses the line information of the statements to identify a read/write on shared variables to insert barriers in the middle, it is unable to do so here since the line numbers will be the same. \autosync ends up in an infinite loop in these scenarios. In contrast, \tool can identify such scenarios since this statement from the source file will be split into two statements, a read followed by a write, in the \boogie representation of the kernel.

\noindent\textbf{Error Parsing vs. SMT Variable Analysis:} \autosync uses regular expressions for parsing the error messages generated by \verifier to identify the locations responsible for causing the data race. This technique makes \autosync extremely fragile to any changes in the output of \verifier. For example, \autosync reports that the program has no errors if, in the output, it does not find texts related to data race or barrier divergence errors. This causes it to misclassify programs with assertion violation errors as error-free programs. \tool relies on the SMT model provided by \verifier to determine which barriers contributed to the error. This approach makes \tool robust and indifferent to the textual output format of \verifier.

\noindent\textbf{Inter-block races:} \tool can propose a fix for inter-block races using CUDA Cooperative Groups for CUDA kernels. No other tool is known to repair inter-block races.

\section{Experiments}
\label{Se:experiments}
In this section, we present our comparison of \tool and \autosync on several CUDA benchmarks. In addition, we present our findings for runs of \tool on OpenCL benchmarks as \autosync cannot tackle OpenCL kernels. The source code of \tool is available at \cite{gpurepairsrc}. The artifacts and the virtual machine used to reproduce the results of this paper are available at \cite{vmcaiartifact} and \cite{vmcaivm}, respectively.

\subsection{Experimental Setup}
\label{Se:experimental_setup}
Since \tool depends on \verifier as an oracle, the implementation of \tool uses the same technology stack as \verifier. The instrumentation and repair stages are built using the .NET Framework with C\# as the programming language. As mentioned in \sectref{workflow}, there are several tools involved in the pipeline of \tool. Specified below are the version numbers of the tools used in the experimentation. We use the \zth solver \cite{de2008z3,bjorner2015nuz} for determining the barrier assignments. The tools used in \tool with their versions are:
 \href{http://llvm.org/svn/llvm-project/llvm/branches/release_60/}{\app{LLVM} 6.0.1},
 \href{http://llvm.org/svn/llvm-project/cfe/branches/release_60/}{\app{Clang} 6.0.1},
 \href{http://llvm.org/svn/llvm-project/libclc/trunk/}{\app{libclc} (Revision 353912)},
 \href{https://github.com/mc-imperial/bugle}{\app{Bugle} (Revision 15df0ed)},
 \href{https://github.com/mc-imperial/gpuverify}{\verifier (Revision d5a90ec)}, and
 \href{https://github.com/Z3Prover/z3/tree/z3-4.6.0}{\zth 4.6.0}.

The experiments were performed on \emp{Standard\_F2s\_v2 Azure\textsuperscript{\textregistered} virtual machine}, which has 2 vCPUs and 4 GiB of memory. More details on the virtual machine can be found at \cite{azuref2sv2}. For the experiments, a total of $748$ kernels ($266$ CUDA and $482$ OpenCL) were considered. This benchmark set is a union of four independent test suites and publicly available \cite{cudatoolkit} sample programs. \tableref{benchmark_summary} summarizes the various sources of the kernels. The average number of lines of code for this benchmark set is $17.51$, and the median is $11$. $14$ kernels have more than $100$ lines of code, and $47$ have more than $50$ lines of code.

\begin{table}[htp]
\caption{Benchmark Summary}
\label{Ta:benchmark_summary}
\begin{center}

\def\arraystretch{1.1}
\setlength\tabcolsep{7pt}

\begin{tabular}{|l|r|}
\hline

\multicolumn{1}{|c|}{\textbf{Source}} & \multicolumn{1}{|c|}{\textbf{Kernels}} \\ \hline \hline
\verifiername Test Suite (Inc. $482$ OpenCL Kernels) \cite{gpuverifytests} & $658$ \\ \hline
NVIDIA GPU Computing SDK v5.0 \cite{cudatoolkit} & $56$ \\ \hline
\autosyncname Micro Benchmarks \cite{anand2018automatic} & $8$ \\ \hline
\toolname Test Suite (Inc. $16$ examples for & \multirow{2}{*}{$26$} \\
CUDA Cooperative Groups) & \\ \hline

\end{tabular}

\end{center}
\end{table}

All the experiments were performed with a timeout of $300$ seconds for each benchmark for each tool. By default, the weight of a grid-level barrier ($gw$) is taken as $12$, and the weight of a barrier inside a loop ($lw$) is $10$. For nested loops, the loop-depth is computed, and the loop weight ($lw$) is raised to the power of the loop-depth.

\subsection{Results}
\label{Se:experimental_results}

\begin{table}[htp]
\caption{Count of kernels grouped by category}
\label{Ta:results}
\centering

\def\arraystretch{1.1}
\setlength\tabcolsep{7pt}

\begin{tabular}{|l|r|r|r|}
\hline

\multicolumn{1}{|c|}{\textbf{Category}} & \multicolumn{1}{|c|}{\textbf{\autosyncname}} & \multicolumn{2}{|c|}{\textbf{\toolname}} \\ \hline
\multicolumn{1}{|c|}{} & \multicolumn{1}{|c|}{\textbf{CUDA}} & \multicolumn{1}{|c|}{\textbf{CUDA}} & \multicolumn{1}{|c|}{\textbf{OpenCL}} \\
\multicolumn{1}{|c|}{Total Benchmarks} & $266$ & $266$ & $482$ \\ \hline \hline
Verified by \verifiername & $152$ & $152$ & $331$ \\ \hline
\quad No changes made by the tool & $146$ & $138$ & $293$ \\
\quad Changes recommended by the tool & $0$ & $\MB{13}$ & $25$ \\
\quad Errors  & $6$ & $\MB{0}$ & $10$ \\
\quad Timeouts ($300$ seconds) & $0$ & $1$ & $3$ \\ \hline \hline

Data Race/Barrier Divergence Errors & \multirow{2}{*}{$89$} & \multirow{2}{*}{$89$} & \multirow{2}{*}{$69$} \\
identified by \verifiername & & & \\ \hline
\quad Repaired by the tool & $31$ & $\MB{43}$ & $33$ \\
\quad Repaired using grid-level barriers & $0$ & $\MB{15}$ & $0$ \\
\quad Could not be repaired by the tool & $10$ & $20$ & $34$ \\
\quad Errors  & $14$ & $\MB{0}$ & $0$ \\
\quad Timeouts ($300$ seconds) & $34$ & $\MB{11}$ & $2$ \\ \hline \hline

Unrepairable errors identified by & \multirow{2}{*}{$25$} & \multirow{2}{*}{$25$} & \multirow{2}{*}{$82$} \\
\verifiername & & & \\ \hline
\quad Handled gracefully by the tool & $0$ & $\MB{25}$ & $82$ \\
\quad False Positives & $24$ & $\MB{0}$ & $0$ \\
\quad Errors  & $1$ & $\MB{0}$ & $0$ \\ \hline

\end{tabular}
\end{table}

\tableref{results} summarizes the results obtained from running the benchmark suite with \tool and \autosync. \autosync does not support OpenCL; therefore, results for OpenCL are only applicable for \tool. Numbers in bold indicate better results.

The table categorizes the results into three categories based on the output of \verifier. The first category includes all the kernels for which \verifier concluded that there were no errors. For all $152$ CUDA kernels that fall in this category, \autosync crashed in $6$ of these benchmarks, and even though \verifier did not give any errors, \tool suggested removal of unnecessary barriers for $13$ CUDA kernels and $25$ OpenCL kernels. Removal of unnecessary barriers is a feature exclusive to \tool. \autosync attempts only to insert barriers so as to avoid data race errors.

The second category includes the kernels for which \verifier had identified data races or barrier divergence errors. For $10$ benchmarks, \autosync stated that the error could not be repaired. Out of these $10$, \tool was able to repair $6$, it timed out for $2$, and it could not repair $2$ of these. The final category involves the kernels that had either assertion errors or errors thrown by \app{Clang} or \app{Bugle} or had invalid invariants. Repairing these kernels is beyond the scope of either \autosync or \tool. \autosync claimed that a solution was found for $24$ out of the $25$ CUDA kernels in this category, but those were found to be false positives. \autosync checks the textual error messages for any information related to data races and barrier divergence, and if it does not find anything, \autosync treats it as a success. In contrast, \tool returned the same error code as \verifier for all the benchmarks in this category. This category highlights the fragile nature of \autosync, as it syntactically depends on the output of \verifier.

It is evident that \tool provides much more coverage as it can handle OpenCL kernels. Even for CUDA kernels, \tool provides better coverage as it can repair more programs, support inter-block synchronization using CUDA Cooperative Groups, and exits gracefully for a larger number of kernels.

\begin{figure}[htp]
\centering

\begin{minipage}{.45\textwidth}
    \centering
    \begin{tikzpicture}[scale=0.65]
    
    \pgfplotstableread{figures/data/time_scatter.dat}{\data}
    \selectcolormodel{gray}
    
    \begin{axis}[
        xlabel={GPURepair},
        ylabel={AutoSync},
        scaled ticks = false,
        tick label style={
            /pgf/number format/fixed,
            /pgf/number format/precision=3
        },
        xmin=1,
        xmax=300,
        ymin=1,
        ymax=300,
        xmode=log,
        ymode=log,
        log ticks with fixed point
    ]
    
    \addplot[
        scatter=true,
        only marks,
        mark=*,
        scatter src=explicit symbolic,
        scatter/classes={
            a={mark=*}
        }
    ]
    table[x=gpurepair,y=autosync,meta=label]{\data};
    \addplot [black,samples at={0,1}] {x};
    \draw [black,solid] (rel axis cs:0,0) -- (rel axis cs:1,1);
    
    \end{axis}
    \end{tikzpicture}
    \caption{Runtime in seconds (Log Scale)}
    \label{Fi:time_scatter_log}
\end{minipage}
\hskip 10pt
\begin{minipage}{.45\textwidth}
    \centering
    \begin{tikzpicture}[scale=0.65]
    
    \pgfplotstableread{figures/data/time_repaired.dat}{\data}
    \selectcolormodel{gray}
    
    \begin{axis}[
        xlabel={GPURepair},
        ylabel={AutoSync},
        scaled ticks = false,
        tick label style={
            /pgf/number format/fixed,
            /pgf/number format/precision=3
        },
        xmin=1,
        xmax=300,
        ymin=1,
        ymax=300,
        xmode=log,
        ymode=log,
        log ticks with fixed point
    ]
    
    \addplot[
        scatter=true,
        only marks,
        mark=*,
        scatter src=explicit symbolic,
        scatter/classes={
            a={mark=*}
        }
    ]
    table[x=gpurepair,y=autosync,meta=label]{\data};
    \addplot [black,samples at={0,1}] {x};
    \draw [black,solid] (rel axis cs:0,0) -- (rel axis cs:1,1);
    
    \end{axis}
    \end{tikzpicture}
    \caption{Runtime in seconds - Repair Candidates (Log Scale)}
    \label{Fi:time_repair_log}
\end{minipage}

\end{figure}

As described in \tableref{results}, there were $25$ benchmarks in the third category where \verifier itself throws an irreparable error (e.g., either an assertion violation or errors thrown by other stages). We provide a time comparison in the form of a scatter plot shown in \figref{time_scatter_log} for the remaining $241$ CUDA benchmarks, which are either certified as error-free by \verifier or contain data race and/or barrier divergence errors. Each benchmark has been executed $3$ times for each of the tools, and the average time for these $3$ runs is taken into consideration. We used the average since there was a negligible difference between the median and the average.

Out of the $241$ benchmarks, \autosync was faster in $178$ cases, whereas \tool was faster for $63$ benchmarks. \verifier did not show any error for $152$ out of these $241$ benchmarks. For these benchmarks, \autosync did not have to put any further efforts. In contrast, \tool attempts to determine if there are any programmer-inserted barriers that are unnecessary and could be removed. This explains \tool being slower for some of the benchmarks. \figref{time_repair_log} shows a run time comparison for $89$ benchmarks for which \verifier found data race/barrier divergence errors. \tool performs significantly better than \autosync when a kernel requires repair. This is evident from \figref{time_repair_log} as \autosync was faster on $32$ benchmarks, whereas \tool was faster on $57$ benchmarks out of these $89$ benchmarks. Note that if any of the tools crash on a benchmark, we consider that run to have timed out; that is, as a benchmark run that took 300 seconds.

\begin{table}[htp]
\caption{Comparison of \autosync and the various configurations of \tool}
\label{Ta:configuration_comparison}
\centering

\def\arraystretch{1.1}
\setlength\tabcolsep{7pt}

\begin{tabular}{|l|r|r|r|r|r|}
\hline

\multicolumn{1}{|c|}{} & \multicolumn{2}{|c|}{\textbf{All Kernels}} & \multicolumn{3}{|c|}{\textbf{Repaired $+$ Unchanged}} \\
\multicolumn{1}{|c|}{} & \multicolumn{2}{|c|}{($241$)} & \multicolumn{3}{|c|}{\textbf{($28+133=161$)}} \\ \hline
\multicolumn{1}{|c|}{\textbf{Tool}} & \multicolumn{1}{|c|}{\textbf{Total}} & \multicolumn{1}{|c|}{\textbf{Median}} & \multicolumn{1}{|c|}{\textbf{Total}} & \multicolumn{1}{|c|}{\textbf{Median}} & \multicolumn{1}{|c|}{\textbf{Verifier}} \\
\multicolumn{1}{|c|}{\textbf{(Configuration)}} & \multicolumn{1}{|c|}{\textbf{Time}} & \multicolumn{1}{|c|}{\textbf{Time}} & \multicolumn{1}{|c|}{\textbf{Time}} & \multicolumn{1}{|c|}{\textbf{Time}} & \multicolumn{1}{|c|}{\textbf{Calls}} \\
\multicolumn{1}{|c|}{} & \multicolumn{4}{|c|}{\textbf{(in seconds)}} & \multicolumn{1}{|c|}{} \\ \hline \hline
\autosync & $17076$ & $1.43$ & $384$ & $1.24$ & $216$ \\ \hline
\tool & $5810$ & $1.76$ & $823$ & $1.57$ & $271$ \\ \hline
\tool \xspace \TT{--maxsat} & $5940$ & $1.75$ & $887$ & $1.54$ & $306$ \\ \hline
\tool \xspace \TT{--disable-grid} & $4450$ & $1.72$ & $660$ & $1.56$ & $254$ \\ \hline
\tool & \multirow{2}{*}{$5430$} & \multirow{2}{*}{$1.81$} & \multirow{2}{*}{$754$} & \multirow{2}{*}{$1.56$} & \multirow{2}{*}{$250$} \\
\quad \TT{--disable-inspect} & & & & & \\ \hline
\tool \xspace \TT{--disable-grid} & \multirow{2}{*}{$4225$} & \multirow{2}{*}{$1.79$} & \multirow{2}{*}{$621$} & \multirow{2}{*}{$1.51$} & \multirow{2}{*}{$235$} \\
\quad \TT{--disable-inspect} & & & & & \\ \hline

\end{tabular}
\end{table}

The default configuration of \tool uses the $mhs$ strategy to solve the clauses, enables exploring grid-level barriers to find a solution, and inspects pre-existing barriers for removal if deemed unnecessary. The solver type can be changed to $MaxSAT$, and the usage of grid-level barriers and inspection of pre-existing barriers can be disabled through command-line options. In addition, the weight of grid-level barriers and the weight of barriers nested within loops can also be overridden using command-line options. \ref{Apx:running} gives a brief description on how to run \tool; detailed documentation can be found at \cite{gpurepairsrc}. \tableref{configuration_comparison} compares the time taken by \autosync and \tool on different configurations. The total time taken by \autosync is higher, primarily because $300$ seconds were counted every time any of the tools crashed. This analysis demonstrates that \tool is more robust than \autosync while performing quite close to \autosync with respect to runtime.

Additional experiments and analysis are included in \ref{Apx:results}, which provides more experimental details of the various configurations of \tool(\ref{Apx:configcomp}), the code size of the benchmarks (\ref{Apx:codesize}), analysis of the errors encountered by \autosync (\ref{Apx:erroranalysis}), case studies (\ref{Apx:examples}), and how to run \tool (\ref{Apx:running}).

\section{Conclusion}
\label{Se:conclusion}
This tool paper introduces \tool, which can fix barrier divergence errors and remove unnecessary barriers in addition to \textsf{\autosyncname's} ability to fix data races. \tool has the additional capability to handle both CUDA and OpenCL kernels. Most importantly, \tool has a unique feature for suggesting a fix for inter-block races using Cooperative Groups in CUDA.

With extensive experimental evaluation on the benchmark suites (consisting of 700+ CUDA/OpenCL kernels), we have affirmed the superiority of our work. Our experimental results clearly show that \tool provides far more coverage than \autosync. For $65\%$ of the total benchmarks, \autosync was not applicable as $482$ out of the $748$ benchmarks were OpenCL kernels. Even for CUDA kernels, \tool is able to provide more coverage and is able to repair more kernels. \tool is also faster than \autosync when a kernel indeed requires repair.

\section*{Acknowledgements}
We thank the anonymous reviewers for their helpful comments and the authors of \autosync for providing the source-code under a public license. We also thank the Ministry of Education, India, for financial support.

\bibliographystyle{splncs04}
\bibliography{references}

\appendix
\renewcommand{\thesection}{Appendix \Alph{section}}
\renewcommand{\thesubsection}{\Alph{section}.\arabic{subsection}}
\section{More Experiments and Results}
\label{Apx:results}

\subsection{Configuration Comparison}
\label{Apx:configcomp}
As mentioned in \sectref{repair_algorithm}, \tool computes the minimal-hitting-set ($mhs$) of the clauses obtained in each run of the algorithm. Using $mhs$ can avoid expensive calls to the $MaxSAT$ solver at every iteration. In principle, using the $mhs$ strategy should always outperform the $MaxSAT$ strategy if the same set of error traces are seen in the same sequence. However, since the approach with/without $mhs$ would result in different sets of traces being witnessed and in different sequences, $mhs$ may not always outperform on every benchmark. For example, consider that in the first iteration, we obtain a clause $a \vee b$ from GPUVerify. $mhs$ could choose $a$ to be the solution, and $MaxSAT$ could choose $b$ to be the solution. Both of the solutions are valid for the clause, but choosing $b$ could fix the kernel, and choosing $a$ may not, thus forcing more iterations. These ``lucky'' choices can introduce deviations regarding which solver would perform better on different benchmarks. \figref{mhs_maxsat} shows the runtime comparison of these solvers with a $300$ seconds timeout for each benchmark.

The $mhs$ (default) strategy of \tool takes lesser time to run all the $748$ benchmarks compared to the $MaxSAT$ strategy, taking a total time of $9145$ seconds. When grid-level barriers and inspection of programmer-inserted barriers are disabled, the time reduces to $6467$ seconds, but this comes at the cost of not repairing a few kernels that require grid-level synchronization and not optimizing any.

The $mhs$ strategy of \tool made a total of $415$ calls to \zth. Note that these do not include the calls made to \zth by \verifier. There were $9$ kernels for which $mhs$ could not solve the input clauses (a possibility explained in~\sectref{repair_algorithm}), and the $MaxSAT$ solver was used as a fallback. In all of these cases, the $MaxSAT$ solver also returned UNSAT.

\begin{figure}[htp]
\centering

\begin{minipage}{.45\textwidth}
    \centering
    \begin{tikzpicture}[scale=0.65]
    
    \pgfplotstableread{figures/data/time_mhs_maxsat.dat}{\data}
    \selectcolormodel{gray}
    
    \begin{axis}[
        xlabel={GPURepair (mhs)},
        ylabel={GPURepair (MaxSAT)},
        scaled ticks = false,
        tick label style={
            /pgf/number format/fixed,
            /pgf/number format/precision=3
        },
        xmin=1,
        xmax=300,
        ymin=1,
        ymax=300,
        xmode=log,
        ymode=log,
        log ticks with fixed point
    ]
    
    \addplot[
        scatter=true,
        only marks,
        mark=*,
        scatter src=explicit symbolic,
        scatter/classes={
            a={mark=*}
        }
    ]
    table[x=mhs,y=maxsat,meta=label]{\data};
    \addplot [black,samples at={0,1}] {x};
    \draw [black,solid] (rel axis cs:0,0) -- (rel axis cs:1,1);
    
    \end{axis}
    \end{tikzpicture}
    \caption{Runtime in seconds - mhs vs. MaxSAT (Log Scale)}
    \label{Fi:mhs_maxsat}
\end{minipage}
\hfil
\begin{minipage}{.5\textwidth}
    \centering
    \begin{tikzpicture}[scale=0.7]
    
    \selectcolormodel{gray}
    
    \begin{axis}[
        xbar, xmin=0, xmax=450,
        xlabel={Kernel Count},
        symbolic y coords={
            {$> 100$},
            {$76-100$},
            {$51-75$},
            {$26-50$},
            {$11-25$},
            {$<= 10$}
        },
        ytick=data,
        ylabel={Lines of Code},
        nodes near coords,
        nodes near coords align={horizontal},
        height=200pt, width=200pt
    ]
    
    \addplot coordinates {
        (14,{$> 100$})
        (9,{$76-100$})
        (24,{$51-75$})
        (38,{$26-50$})
        (293,{$11-25$})
        (370,{$<= 10$})
    };
    
    \end{axis}
    \end{tikzpicture}
    
    \caption{Lines of Code}
    \label{Fi:code_lines}
\end{minipage}

\end{figure}

\subsection{Source Code Size}
\label{Apx:codesize}

\figref{code_lines} shows the size of the kernels in terms of lines of code. The average number of lines of code for the $748$ kernels in the evaluation set was $17.51$, and the median was $11$. $14$ kernels had more than $100$ lines of code, and $47$ had more than $50$ lines of code. The majority of the kernels had less than $25$ lines of code. The kernel with the highest number of lines of code had $639$ lines.

The instrumentation step of \tool happens on the \boogie code and not on the source code. A line of source code could result in zero (e.g., code comments) or more \boogie commands. \figref{boogie_commands} shows the size of the kernels in terms of \boogie commands. The average number of \boogie commands for the $734$ kernels for which \boogie code was generated was $25.72$, and the median was $11$. The majority of the kernels had less than $25$ commands. The kernel with the highest number of \boogie commands had $1793$ commands.

\begin{figure}[htp]
\centering

\begin{minipage}{.45\textwidth}
    \centering
    \begin{tikzpicture}[scale=0.7]
    
    \selectcolormodel{gray}
    
    \begin{axis}[
        xbar, xmin=0, xmax=450,
        xlabel={Kernel Count},
        symbolic y coords={
            {$> 100$},
            {$76-100$},
            {$51-75$},
            {$26-50$},
            {$11-25$},
            {$<= 10$}
        },
        ytick=data,
        ylabel={Boogie Commands},
        nodes near coords,
        nodes near coords align={horizontal},
        height=200pt, width=200pt
    ]
    
    \addplot coordinates {
        (28,{$> 100$})
        (12,{$76-100$})
        (38,{$51-75$})
        (99,{$26-50$})
        (192,{$11-25$})
        (379,{$<= 10$})
    };
    
    \end{axis}
    \end{tikzpicture}
    
    \caption{Boogie Commands}
    \label{Fi:boogie_commands}
\end{minipage}
\hfil
\begin{minipage}{.5\textwidth}
    \centering
    \begin{tikzpicture}[scale=0.7]
    
    \selectcolormodel{gray}
    
    \begin{axis}[
        xbar, xmin=0, xmax=300,
        xlabel={Kernel Count},
        symbolic y coords={
            {$> 100$},
            {$21-100$},
            {$11-20$},
            {$6-10$},
            {$4-5$},
            $3$,
            $2$,
            $1$,
            $0$
        },
        ytick=data,
        ylabel={Barrier Count},
        nodes near coords,
        nodes near coords align={horizontal},
        height=200pt, width=200pt
    ]
    
    \addplot coordinates {
        (5,{$> 100$})
        (59,{$21-100$})
        (61,{$11-20$})
        (110,{$6-10$})
        (92,{$4-5$})
        (50,$3$)
        (95,$2$)
        (59,$1$)
        (203,$0$)
    };
    
    \end{axis}
    \end{tikzpicture}
    
    \caption{Kernels with 'n' Instrumented Barriers}
    \label{Fi:kernels_barriers}
\end{minipage}

\end{figure}

The repair step of \tool depends on the number of barriers that are instrumented in code, which depend on the number of shared variables and how many times they have been used.

\figref{kernels_barriers} shows the number of barrier variables introduced in the instrumentation stage of \tool. This also includes the barrier variables introduced for existing barriers provided by the programmer. Out of the $734$ kernels that reached the instrumentation stage (inclusive of CUDA and OpenCL kernels), more than $50\%$ had less than three instrumented barriers introduced during the instrumentation stage, and there were $5$ kernels with more than $100$ instrumented barriers. The highest number of barriers, $588$, were introduced in the \path{OpenCL/vectortests/float4initialisation} test-case, followed by \path{CUDASamples/6_Advanced_shfl_scan_shfl_intimage_rows}, which had $356$ instrumented barriers. For all the kernels, the time taken by the instrumentation stage was less than a second.

\subsection{\autosyncname Errors}
\label{Apx:erroranalysis}
\autosync crashes for $21$ kernels. The root cause of these crashes can be classified into two broad categories - error message parsing and source code parsing.

$8$ kernels crash because of error message parsing. These kernels have a method named ``race'' that causes the problem. \verifier throws a warning about aliasing, which mentions the method name, and \autosync mistakes it for an error related to data race because of the regular expressions that are used for error parsing. For $3$ kernels, changing the method name worked, and \autosync was able to repair them. The solution for $3$ of these kernels involved inter-block synchronization, which \autosync does not support. For the remaining $2$, \autosync timed out.

The remaining $13$ kernels crash because of source code parsing. \autosync has a component that initially parses the source code, which is used at a later stage to decide where a barrier can be inserted. $6$ kernels crashed because the if-else statements had the opening brace in the same line as the if statement. After moving the brace to a new line, \autosync was able to work for $5$ of the kernels. For the sixth kernel, \autosync returned a false-positive result since the kernel had a Bugle error. The remaining $7$ kernels crashed because of complicated source code parsing issues in \autosync. We were not able to find a fix in the \autosync code for these. This component ideally would need to be a full-fledged C parser, which is quite cumbersome to build.

\subsection{Case Studies}
\label{Apx:examples}

In this section, we showcase some of the examples that needed special consideration while designing \tool. Consider \coderef{unrepairable}; there is a read-write race between \codeline{un:read} and \codeline{un:write}. The repair algorithm identifies this and inserts a barrier before \codeline{un:write}. However, since these statements are inside an if block, adding the barrier leads to a barrier divergence. The conflict between having the barrier and encountering a divergence error and not having it and encountering a race causes \tool to consider it an unrepairable kernel. Neither \tool nor \autosync can repair such kernels.

\coderef{single_line_race} showcases an example of a shared array being read and written to in the same statement. When \app{Bugle} transforms this kernel to \boogie, it breaks these statements into two statements, as shown below. \tool leverages this to recommend a barrier placement between the read and write to fix the kernel. As mentioned in \sectref{comparison}, \autosync cannot handle such scenarios.

\begin{center}
\begin{tabular}{lcl}
	\TT{A[tid] = A[tid + 1];} & $\quad\longmapsto\quad$ & \TT{int temp = A[tid + 1];} \\
	& & \TT{A[tid] = temp;} \\
\end{tabular}
\end{center}

One of the scenarios that we came across in the \verifier test suite was a write-write race by multiple threads on the same shared array at the same index. This is shown in \coderef{write_write_race}. There is no solution for such kernels, and neither \tool nor \autosync can fix them.

\noindent\begin{tabular}{ll}
	\begin{minipage}{0.45\textwidth}
\begin{lstlisting}[caption=Unrepairable Kernel, label={Cd:unrepairable}, captionpos=b, xleftmargin=2em, escapechar=\^]
__global__ void race (int* A) {
  int tid = threadIdx.x;
  if (tid % 2 == 0)
  {
    int temp = A[tid + 1]; ^\label{Line:un:read}^
	A[tid] = temp; ^\label{Line:un:write}^
  }
}
\end{lstlisting}

	\end{minipage} &
	\begin{minipage}{0.55\textwidth}
\begin{lstlisting}[caption=Single Line Race, label={Cd:single_line_race}, captionpos=b, xleftmargin=3em, escapechar=\^]
__global__ void race (int* A) {
  int tid = threadIdx.x;
  A[tid] = A[tid + 1];
}
\end{lstlisting}

\begin{lstlisting}[caption=Write-Write Race, label={Cd:write_write_race}, captionpos=b, xleftmargin=3em, escapechar=\^]
__global__ void race (int* A) {
  A[0] = threadIdx.x;
}
\end{lstlisting}

	\end{minipage}
\end{tabular}

Consider the kernel in \coderef{interfunction_race}. When function inlining is enabled, \verifier can accurately identify that \codeline{ifr:read} and \codeline{ifr:write} cause a read-write race through the functions invoked at these lines of code. It reports the line information by specifying that the lines inside the functions cause the read-write race and also explicitly specifies the lines from where these functions are invoked. \autosync, however, does not process this information correctly and ends up with a code error. \tool takes these cases into account and can place the barrier precisely between the function invocations.

\noindent\begin{tabular}{ll}
	\begin{minipage}{0.45\textwidth}
\begin{lstlisting}[caption=Inter Function Race, label={Cd:interfunction_race}, captionpos=b, xleftmargin=2em, escapechar=\^]
__device__ void write(int* A,
    int tid, int temp) {
  ^\highlightcode{0.55}^A[tid] = temp;
}
__device__ int read(int* A,
    int tid) {
  ^\highlightcode{0.55}^return A[tid+1];
}

__global__ void race (int* A) {
 int tid = threadIdx.x;
 int temp = read(A, tid); ^\label{Line:ifr:read}^
 write(A, tid, temp); ^\label{Line:ifr:write}^
}
\end{lstlisting}

	\end{minipage} &
	\begin{minipage}{0.55\textwidth}
\begin{lstlisting}[caption=Assertion Errors after Adding Barriers, label={Cd:assertion_error}, captionpos=b, xleftmargin=3em, escapechar=\^]
typedef int(*funcType)(int);
__device__ int multiply(int i)
  { return i*2; }
__device__ int divide(int i)
  { return i/2; }

__global__ void race(int *A) {
  int tid = threadIdx.x;
  funcType f = (tid != 0) ?
    multiply : divide;
  A[tid] = (*f)(tid);
  ^\highlightcode{0.5}^__syncthreads(); ^\label{Line:ae:extrabarrier}^
  __assert(__implies(tid != 0,
    A[tid] == 2*tid));
}
\end{lstlisting}

	\end{minipage}
\end{tabular}

Consider the kernel in \coderef{assertion_error}. Without the barrier at the highlighted line, \verifier can successfully verify and conclude that there are no errors. However, when the barrier is introduced at \codeline{ae:extrabarrier}, \verifier reports an assertion error. This false positive is due to the way \verifier abstracts shared state when it encounters a barrier in such cases. This happens even when the barrier is an instrumented barrier. Due to scenarios like these, sometimes \tool is unable to obtain a solution even though one exists. \autosync also needs the verifier to be sound and complete for it to find a solution.

\subsection{Running \toolname}
\label{Apx:running}

\begin{figure}[htp]
\centering
\includegraphics[scale=0.5]{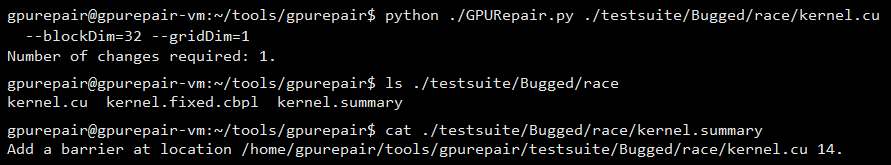}
\caption{\toolname Commands}
\label{Fi:tool_commands}
\end{figure}

As emphasized in \figref{repairer_workflow}, \tool uses multiple tools in its repair process. We provide a python script, which takes care of calling all of the tools in the toolchain in the appropriate order. Once the kernel is repaired, \tool prints the needed changes in the source code and provides a fixed \boogie program as evidence. \figref{tool_commands} shows a sample run and the changes recommended by \tool for this kernel.

\begin{figure}[htp]
\centering
\includegraphics[scale=0.6]{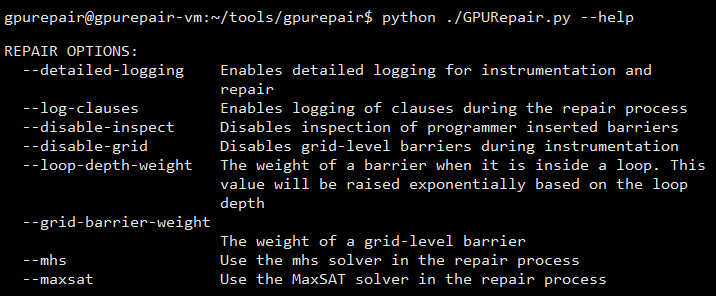}
\caption{\toolname Repair Options}
\label{Fi:tool_options}
\end{figure}

\verifier supports several verification options. These command-line options are documented at \cite{gpuverifyoptions}. All of these can be used with \tool as well. \tool passes these command-line options, if specified during its invocation, to \verifier for the verification process. Along with these, \tool also provides specific options for the repair process. A description of these can be accessed from the help section. \figref{tool_options} shows this.

\end{document}